\documentstyle[12pt]{article}    

\def\be{\begin{equation}}
\def\ee{\end{equation}}
\def\beq{\begin{eqnarray}}
\def\eeq{\end{eqnarray}}

\def\bay{\begin{array}}
\def\eay{\end{array}}

\begin{document}

\begin{titlepage}

\title{On the quantum theory of pure gravitation}
\author{
B. Ram
\\
\\
\footnote{correspondence address}
{\footnotesize
Physics Department, New Mexico State University, Las Cruces, NM 88003, USA}
\\ 
{\footnotesize Umrao Institute of Fundamental Research, A2/214 Janak Puri, New Delhi, 110058, India}
}
\maketitle
\begin{abstract}
A quantum theory of the region of pure gravitation was given earlier in two papers [gr-qc/9908036 (Phys. Lett. A {\bf {265}}, 1 (2000)); gr-qc/0101056]. In this paper I provide further insight into the physics of this region.
\end{abstract}
\end{titlepage}
\begin{center}
{\small ... {\em the solutions which the general theory of relativity allows...partition the three dimensional space into two regions: an inner region, bounded by a smooth two-dimensional null surface, which (the inner region) is incommunicable to the outer region which is, in turn, asymptotically flat.}}
\end{center}
\hspace{8cm} {\small S. Chandrasekhar - Truth and Beauty}\\
\vspace{2cm}

A quantum theory of precisely the inner region, the region of pure gravitation, a quantum not classical region was recently given in two papers \cite{ram1, ram2}. The main aim here is to provide further insight into the physics of this region in order to help other researchers in their quest. 

[The writing of the present paper, the last of the trilogy, is animated by the very recent review paper by Carlip \cite{carlip} which came to my attention. The abstract of Carlip's paper reads:\\
"The problem of reconciling general relativity and quantum theory has fascinated and bedeviled physicists for more than 70 years. Despite recent progress in string theory and loop quantum gravity, a complete solution remains out of reach."]

When I gave a copy of the first paper \cite{ram1} to a bright graduate student in string theory, he came back and told me that he does not agree with me because equation (1), namely
\be \label{r1}
\frac{1}{2}{\dot r}^2 + \frac{L^2}{2r^2} - \frac{m}{r} - \frac{mL^2}{r^3} = \frac{1}{2}(E^2 - 1)\ee
for the test particle does not take into account the back reaction of the test particle. I asked him if he would be satisfied if we set the attributes of the test particle, $E$and $L$, to zero in (\ref{r1}). He said, yes. So I asked him to go home and read the paper again, this time a little more carefully. I took it that he must have realized that the final equation (8) which {\em is}
\be \label{r2}
\left(-\frac{1}{2}\frac{d^2}{dr^2} - \frac{\mu/4}{r}\right)U = -\frac{1}{2}U,\ee
was devoid of the test particle's attributes and hence its back reaction. Equation (2) is the quantum equation of the inner region, the region of pure gravitation.

Putting $E$ and $L$ equal to zero in the classical time-like geodesic equation (\ref{r1}) and then applying the Schr${\ddot o}$dinger prescription is equivalent to the method given in \cite{ram1}. It not only removes the logical inconsistency pointed out by the graduate student, but also makes the value of the Schwarzschild coordinate $r$ that of the boundary of the inner and outer regions i.e. of the horizon with ${\dot r} = 0$ there. This can be seen as follows \cite{six}. Consider the radial geodesic with zero angular momentum \cite{seven}:
\be \label{r3}
\frac{1}{2}{\dot r}^2 - \frac{m}{r} = \frac{1}{2}(E^2 - 1).\ee
Now, using the test particle language, if a test particle in the outer region starts from rest ($\frac{dr}{d\tau} = {\dot r} = 0$) at some finite distance $r_i$ and falls towards the inner region, the relation between the constant $E$, which is usually interpreted \cite{five} as the total energy of the test particle per unit mass \cite{eight}, and the starting distance $r_i$ is given by
\be \label{r5}
E^2 = 1 - 2\frac{m}{r_i}.\ee
It is easy to see from (\ref{r5}) that when $r_i = \infty, E = 1$, as it should be \cite{nine}; and when 
\be \label{r6}
r_i = 2m, \hspace{2cm} E = 0.\ee
One usually interprets \cite{ten} the result (\ref{r6}) by saying that a test particle cannot be at rest at $r = 2m$, or that no material test particle can have ${\dot r} = 0$ at $r = 2m$, the location of the horizon.
However, the more viable interpretation\cite{eleven} is the obvious one, namely, that one {\em can} have ${\dot r} = 0$ at $r = 2m$ with {\em no} test particle, $r$ simply being the position in the ($r, \tau$) space. Thence follows the dynamical equation for the inner ($r, \tau$) region, namely 
\be \label{r7}
\frac{1}{2}{\dot r}^2 - \frac{m}{r} = -\frac{1}{2}.\ee
However the full force of (\ref{r7}) is realized only in its quantized form (2) which is also obtainable by using Schr{$\ddot o$}dinger's prescription in (\ref{r7}).

The quantum equation (2) for the inner region of pure gravitation can also be obtained without employing the notion of a test particle. Consider the Schwarzschild spacetime ($r, t$) represented by the metric \cite{seven}
\be \label{r8}
ds^2 = -\left(1 - \frac{2m}{r}\right) dt^2 + \left(1 - \frac{2m}{r}\right)^{-1}dr^2.\ee
It is static in the outer region $r > 2m$, $r = 2m$ being the static limit. Taking $ds^2 = -d\tau^2$ and dividing by 
$d\tau^2$, one gets
\be \label{r9}
1 = \left(1 - \frac{2m}{r}\right){\dot t}^2 - \left(1 - \frac{2m}{r}\right)^{-1}{\dot r}^2,\ee
or
\be \label{r10}
 1 - \frac{2m}{r} = \left(1 - \frac{2m}{r}\right)^2{\dot t}^2 - {\dot r}^2,\ee
as the equation for the time-like curve in the $(r, t)$ plane. In the outer region $(r \geq 2m)$, $r$ is spacelike and $t$ is timelike. The curve (\ref{r10}) is traceable by keeping ${\dot r} = 0$ from $r = \infty$ to $r = 2m$. At $r = \infty$, ${\dot t} = 1$ and at $r = 2m$, ${\dot t} = \infty$, that is at $r = 2m$ the $t$-time exhausts its full extent by becoming infinite. On the other hand, in the inner region $(r \leq 2m)$, $r$ becomes time-like, i.e. dynamic \cite{twelve}, the equation describing the dynamics is given by
\be \label{r11}
1 - \frac{2m}{r} = -{\dot r}^2\ee
with $r = 2m$ being the {\em dynamic} limit. Equation (\ref{r11}) is (\ref{r7}) and as stated earlier, application of the Schr${\ddot o}$dinger recipe to (\ref{r7}) gives the quantum equation (2) for the inner region of pure gravitation.

I emphasize here that the numerical 1 on the LHS of equation (\ref{r11}) and the $-\frac{1}{2}$ on the RHS of (\ref{r7}) is a manifestation of the fact that $d\tau^2$ maintains its invariance along the time-like curve. In other words, the appearance of $-\frac{1}{2}$ on the RHS of (\ref{r7}) is a direct consequence of general relativity. Its full importance is realized in the quantum equation (2). Equation (2) says that the inner region of pure gravitation is describable, quantum mechanically, as a {\em bound state} with binding energy equal to $\frac{1}{2}$. It has been shown in paper \cite{ram1} that this is possible for infinitely many discrete values of the mass $\mu$, which are given by
\be \label{r12}
\mu_n = 2(n + 1)\omega, \hspace{1cm}n = 0, 1, 2,...\ee
with $\omega = 2$. Equation (\ref{r12}) is a direct result of the fact that equation (2) is equivalent to a four-dimensional quantum oscillator of angular momentum precisely equal to zero, in full agreement with the fact that the classical Schwarzschild metric is angular momentum $a = 0$ case of the classical Kerr metric. Extension of the above quantization procedure to the Kerr inner region leads naturally to the quantization of the angular momentum $a$ in the Kerr metric, equation (2) being generalized to equation (30) of reference \cite{ram2} and the mass $\mu$ to 
\be \label{r13}
\mu_{n, j} = 2(n + j + 1)\omega, \hspace{1cm} n = 0, 1, 2....;j = 0, 1, 2,....n\ee
In equation (12) $2j$ is the angular momentum quantum number of the four-dimensional quantum oscillator, and $j(j + 1)$ corresponds to $a^2$ \cite{thirt}. Modern interpretation of equation (12) automatically leads to the concept of a quantum of pure gravitation of energy  {\em twice} the planck energy and spin angular momentum {\em one}. It [quantum of pure gravitation] has only two states of polarization. It is well known that pure radiation also behaves as a quantum oscillator \cite{fourt}. Thus in that sense, the quantum of pure gravitation is like the quantum of pure radiation (photon), except that the quanta of pure gravitation always come in {\em pairs} \cite{fift}.
At the fundamental level, the pair has spin zero or spin two. (One may be tempted to identify the spin zero pair with the Higgs particle and the spin two pair with the "graviton", but that would make both of them out of experimental reach for at least a century!) Consequently the physical entropy of a large black hole of mass $M$ can be calculated using the same method given by Bose for blackbody radiation \cite{ram1, ram2}. The calculations give the same result as expressed by the Bekenstein-Hawking formula, except that the physical temperature of the black hole is defined by \cite{ram2}
\be \label{temp}
T = \frac{k}{2\pi(1 - \frac{a^2}{M^2})^{1/2}} = \frac{k'}{2\pi}\ee
which, of course, reduces to $\frac{k}{2\pi}$ for a non-rotating black hole. Relation (\ref{temp}) not only gives the correct entropy, $\frac{A}{4}$, for the non-extremal case
 $( a < M)$ but also for the extremal case $(a = M)$, in which case $T = \frac{1}{4\pi M}$ and $A = 8\pi M^2$. It would be interesting to obtain (\ref{temp}) using semiclassical methods.

It is generally thought by relativists that in qunatizing general relativity there is a $t$-time
problem \cite{sixt}. Note however that there is no such problem in the present theory. As emphasized earlier, the inner region is a $(r, \tau)$ region, not a $(r, t)$ region, and the classical dynamical equation (\ref{r7}) involves derivatives only with respect to $\tau$, the proper time. Furthermore, since $d\tau$ is invariant under a general transformation of coordinates, $\tau$ has the same character as that of Newton's 'absolute' time, and, as is clear from (\ref{r7}), the variable $r$ behaves like the Cartesian coordinate of Newtonian mechanics. Consequently, the Schr${\ddot o}$dinger prescription applied to (\ref{r7}) to obtain (2) is truly valid.

Now the occurrence of a {\em classical} singularity at $r = 0$ is much discussed in the literature \cite{sevent}. But quantum mechanically there is {\em no} singularity at $r = 0$ as can be verified by simply calculating the expectation value of $r$. Using $n = 1$, $l = 0$ coulombic wave function (note that for $n = 1$, $m = 1$ in (7) of \cite{ram1}) the expectation value $<r>$, comes out to be $\frac{3}{2}$. 

By now a thoughtful reader may be asking herself:
"What is the quantum equation for the {\em outer} region?"
Of course, it is equation (5) of \cite{ram1}, namely
\be \label{r88}
\left[-\frac{1}{2r^2}\frac{d}{dr}\left(r^2\frac{d}{dr}\right) + \frac{l(l + 1)}{2r^2} - \frac{m}{r} - \frac{ml(l + 1)}{r^3}\right]R(r) = \frac{1}{2}(E^2 - 1)R(r)\ee
 though it is only approximate, and in it $E$ and $l$ are the energy and angular momentum quantum number, respectively, per unit mass of the test particle. She may want to use (\ref{r88}) to calculate Rutherford scattering \cite{eightt}.

In closing I encourage the curious some to apply the above quantization procedure to the Reissner-Nordstr${\ddot o}$m case to see what quantized values of charge it gives.

\vspace{1cm}

{\Large \bf Acknowledgements}

\vspace{1cm}

I gave lectures on this topic at various universities during the past three years. I thank N. Banerjee for hospitality at Jadavpur University in India, and Prof. A. K. Raychaudhuri for asking penetrating questions. I thank P. Vanhaecke for hospitality at the Universite' de Poitiers in France, and his suggesting of the title 'Quantum Mechanics of Black Holes' for this talk, which was subsequently used by me. I much enjoyed the hospitality of A. Beesham at the University of Zululand in South Africa (SA); of S. D. Maharaj and J. Mckenzie of the University of Natal, SA; of M. Govender of the Durban Institute of Technology, SA and of P. Dunsby at the University of Cape Town, SA.


\begin{thebibliography}{99}
\bibitem{seven}
{\small It suffices to do so because of spherical symmetry}.\\
\bibitem{fift}
{\small Obviously, gravitation is so strong that it polarizes itself; note that $\frac{\mu}{4}$ in (2) is the reduced mass of the two halves $\left(\frac{\mu}{2}, \frac{\mu}{2}\right)$}.\\
\end{thebibliography}
\end{document}